\magnification=1200
\font\bigbf=cmbx10 at 14.4truept
\nopagenumbers

\centerline{\bigbf
A natural phenomenon that may pose a severe aircraft hazard?}
\bigskip
\centerline{\bf Thomas Gold}
\medskip
\centerline{\it
Professor Emeritus, Cornell University,
Ithaca NY 14853; Email: tg21@cornell.edu}

\bigskip

There have been many serious aircraft accidents in recent years that have not
had a satisfactory explanation despite exhaustive researches, and that have
certain features in common. Those features include apparently a situation of
extreme urgency and danger, so that there was no time for the flight crew to
communicate details to the flight controllers; in some cases there were
circumstances that seemed quite unexpected and perplexing to the flight crew,
suggesting an urgent need to override the usual automated control systems and
manually put the plane into a steep dive. In several cases this was followed by
actions to avoid excessive speed that would threaten the structural integrity
of the aircraft. Several accidents have another feature in common: they
occurred along the edge of the North-Eastern American continental shelf. These
include, among others, TW 800 on July 17, 1996, Swissair 111 on September 2,
1998, Egypt
Air 990 on October 1999, and also the crash of J. F. Kennedy Jr. The case of
the EgyptAir crash has recently come under public debate again as some new
information has become known, and the explanation tentatively offered by the
National Transportation Safety Board (NTSB), suggesting a suicide attempt by a
co-pilot, has come under strong attack by Egyptian authorities, and does not
fit with the new information.

In view of the statistically quite improbable occurrence of these accidents, it
seems prudent now to widen the search to causes that have so far not been
included among possible aircraft hazards, and that have possibly a relationship
to geographical features. Among such, the massive emission of gases from the
seafloor (or land surface) seems to us most worthy of attention.

Massive sudden eruptions of gases have occurred in many locations, bursting up through the ground both from ocean floors and from dry land. They often occur repetitively in the same area, and on land create what is known as "mud volcanoes". The amounts of expelled material accumulated in some mud volcanoes in the last million years are as large as 10 or 20 billion tons, and the estimates of the amounts of gas responsible are several times larger than that.
The erupting gases are usually dominated by methane. Since methane is lighter
than air, it races upwards at high speed. Many cases are known where the gas
spontaneously ignited, and flames to a height of 6,000 ft have been
photographed from Baku, in the active mud volcano area on the West shore of the
Caspian Sea. Much higher brief flashes have been reported, up to 30,000 ft but
these were too brief to be photographed. Massive flammable gas eruptions at or
near times of earthquakes (before, during or after) are reported in historical
and in recent times from many parts of the world.

Similar eruptions are indicated on the sea floors, where large areas are
densely covered with "pockmarks", quite characteristic circular features in the
ocean mud, with diameters of between 10 and 200 meters. These features were
first detected in the North Sea by Dr. Martin Hovland, of Statoil, (the
Norwegian oil company), overlying known gas and oil fields. Similar fields have
since been detected in many parts of the world by sonar, again often showing a
relation to underlying hydrocarbon fields, and also there showing features of
repetition of outbursts, with methane again the major component. Both in mud
volcanoes and in
pockmark fields the emitted quantities of gas in any single event may well
amount to some millions of tons.

Another set of observations has now to be added: it is the occurrence of
"mystery clouds" in the air. Satellite photography over a ten year period
revealed more than two hundred clouds that rose up at a high speed from a small
area of land or sea, forming an expanding funnel.  Temperature observations
showed a much lower temperature in the funnel cloud than in the outside air at
the same height, and this implied that the rising gas must be one that is
intrinsically much lighter than air. Only methane and hydrogen are candidates,
and both are combustible. The largest such cloud on record was seen and
reported by several airline pilots flying between Tokyo and Alaska, North-East
of Japan, on April 9, 1984. They described it as a mushroom cloud that reached
up to 50,000 ft, attaining a diameter of more than 200 miles.

Evidence of massive gas emissions have recently been reported by the Woods Hole
Oceanographic Institute, who conducted a sonar survey of the mid-Atlantic US
continental shelf edge. Along a major fault line they found many and very large
pockmarks, similar to those described by Dr. Hovland, indicating that sudden
almost explosive gas eruptions had taken place there. Also recent reports from
the Province of Quebec, of frequent and large displays of lights in the sky,
clearly related to the swarm of earthquakes between November 1988 and end
of January 1989 in the region of Sanguenay and Quebec City, leave little doubt
that massive gas eruptions occurred there, with some flames reaching high into
the sky. Altogether 46 such sightings were recorded in that period, some but
not all coincident with earthquake shocks.  Earthquake-related lights have been
well known and reported since antiquity, and indeed one very large event
involving gas flames was reported in 1663, not far from the Sanguenay region,
close to the St.Lawrence River.

I had investigated in 1982 a "near disaster" of a British Airways 747 plane
flying at 37,000 ft over a volcanic region of Java. All four engines stopped
shortly after it had entered a visible but tenuous volcanic cloud. After
gliding down to 15,000 ft without power, and there apparently leaving the
cloud, all engines could be started again immediately. The same sequence of
events was experienced two weeks later by an Air Singapore 747 plane over a
nearby region, and many years later by a KLM flight over the Aleutian
Islands. A gas lighter than air, and hence combustible, must have been
responsible in all three cases, to have carried small volcanic dust grains to
these altitudes, and its combustion may have been responsible for the engine
failures that were so sharply limited to the flight within the cloud, probably
due to the fuel-rich and oxygen poor mixtures of the gas adding to the airplane
fuel. Gas eruptions of volcanoes are known of either kind: eruptions of a
ground-hugging
heavy gas identified as carbon dioxide, but also eruptions of a light and
flammable gas, probably methane, whose density is a little more than half that
of air.

With three large planes having come so close to disaster, but yet able to give
a precise account of the events, one has to take the threat of gas emission
seriously. The belief that such emissions can come only from volcanoes has been
voiced, but is clearly wrong in view of the facts already mentioned. What
threats would massive gas emissions pose for aircraft?

One effect I have already described: the possibility of inducing failure of all
engines. But several other aircraft hazards have also to be considered. One is
due to the great upward speed the light gases would have, greatly in excess of
the vertical speeds in ordinary atmospheric turbulence, and structural damage
to the plane or serious injuries to persons may result from the ensuing violent
vertical movement. The ignition and explosion of a large mass of gas external
to the plane may be initiated by the engine exhausts and may be violently
destructive, yet the recovered airplane skin would not show the shrapnel holes
that would be the usual signs of explosions.

Other consequences of gas emissions are the dangerous and misleading
indications that the flight instruments would provide. Air speed indicators and
air pressure altimeters would give quite false and fluctuating readings. The
autopilots, programmed for air, may have totally erroneous responses in the
light gas, as indeed may the pilots themselves, who would be perplexed by a
situation they had never encountered or contemplated before.

A further hazard is that clouds of low density gas may not support a plane,
even at a flying speed that would be amply high enough in air. This would cause
a stall of the aircraft, or be preceded by automatic stall-warning that
requires the pilot to turn the nose down into a dive, and then confront the
danger of excessive speed.

Then there are the various fire hazards resulting from combustible air-gas
mixtures, especially in some confined spaces in the airplane where flames could
be supported, even if the same gas-air mixture would readily be extinguished in
the external high speed airflow. That danger may be highest in cable ducts
where damage could destroy the airplane control system.

The North-Eastern coastline or edge of the continental shelf of the US and
Canada, is the northward continuation of the line whose investigation I have
already mentioned. This extension also has a history of earthquakes and gas
emission from sand beaches and water surfaces beyond the shoreline. Such
emissions had not ceased around the times of the aircraft disasters. A large
number of reports were phoned in to police and emergency services in New
Brunswick and Nova Scotia on October 27, about three days before the Egypt Air
crash, stating that at 9:30 p.m. a large fireball had been seen streaking
across the night sky. The details reported did not correspond to a meteorite,
but included reports of flames and events much slower than those caused by
meteors. A peak in the number of reports recorded prior to an event must be
taken seriously, if the number greatly exceeds the number on other days, as was
the case here. There were similar reports also before and after the TW 800
crash. There was also a report from Swiss Air 111 of a strange smell about
three minutes before the declaration of emergency. This is particularly
suggestive of gas effects, as a similar report was made in one of the near
accidents over Java, where gas certainly was involved.

We may then wish to investigate whether some features of aircraft disasters
along this region, the four disasters mentioned and several others that have
also occurred along this corridor, could have an explanation in terms of the
list of hazards I have mentioned, or others that have not yet been considered,
that could be attributed to gas eruptions.

Mr. Jack Reed retired from the Sandia National Laboratory, an expert in sound
propagation, has noted that the "loud" boom heard by many eye witnesses at the
time of the TW 800 crash on a 25 mile stretch of Long Island, nearest point to
the plane 15 miles away, was far too loud to have been caused by the proposed
explosion of the empty central fuel tank. In his view a one ton bomb of TNT
would have been the least required to make such a sound at that distance. Nor
would such
an explosion have caused the various external luminous phenomena that have been
reported by many. Also it is doubtful that an explosion of such a small amount
of fuel vapor could have had the power to tear off the entire front section of
the fuselage. The absence of shrapnel holes in the recovered skin of TW 800 was
taken to exclude a bomb explosion inside or outside the plane. However, a
massive external gas explosion would produce no shrapnel.

The facts newly announced about the EgyptAir disaster make clear that a
deliberate dive had seemed imperative to the pilot then at the controls, and
that a dangerous overspeed situation had then arisen. After a brief recovery to
level flight, again a dive seemed imperative, and the overspeed may then have
destroyed the plane.

There are many steps that can be taken to find whether the sequence of
disasters along this heavily traveled corridor may be due to gas emissions. As
an immediate step I urge the continuation of the sonar search for pockmarks on
the ocean floor along this coastline in the regions of the four disasters
mentioned and others that occurred near this geographical line, since this will
have a good chance of showing whether these accidents were indeed over
locations at which strong gas outbursts had occurred. A routing change may then
be indicated
as the first step to avoid further disasters.

\bye